\documentclass[aps,prl,preprint,showpacs]{revtex4}
\begin{document}
%\draft
\title{
Strong charge transfer effects in the Mg2p$^{-1}$ core-level spectrum of MgB$_2$ 
 }  

\author{
N.V.\ Dobrodey$^a$, A.I.\ Streltsov$^a$, L.S.\ Cederbaum$^a$}
\affiliation{$^a$Theoretische Chemie,
Universit\"at Heidelberg, D-69120 Heidelberg, Germany}
\author{
C.\ Villani$^b$ and F.\ Tarantelli$^b$}
\affiliation{$^b$Dipartimento di Chimica and I.S.T.M.-C.N.R., Universit\`{a} di Perugia, I-06123 Perugia, Italy} 
\date{\today}
%\maketitle

\begin{abstract}
The two available Mg2p$^{-1}$ core-level spectra of the recently discovered high-temperature superconducting
MgB$_2$ crystal exhibit interesting structures, but contradict each other. This motivated us to
perform {\em ab initio\/} calculations on the core-level spectra of cluster models. The
computed spectra reveal unusual rich and intense structures triggered by a B2p$_z\rightarrow$Mg3s,3p
charge transfer.
\end{abstract}

\maketitle
%\twocolumn
%\onecolumn
%\narrowtext

\section{Introduction}

The discovery of superconductivity below 39~K in MgB$_2$ \cite{1} has triggered an
avalanche of publications devoted to this and related compounds. The main
reason for such an unprecedented interest in MgB$_2$ is that the
mechanism responsible for the transition from a normal to a superconducting state
at the remarkably high temperature of 39~K seems to be the conventional BCS
phonon pairing \cite{2,3,4,5,6}. Thus, studying 
properties of MgB$_2$ may give a clue to the problem of the
high-temperature superconductivity, where numerous scenarios other than phonon
pairing have been proposed (see Ref.~\cite{7}).

In contrast to the known high-temperature superconductors that contain atoms
of $d$ and $f$ elements, MgB$_2$ consists of light atoms, which allows one to
treat this compound using accurate 
{\em ab initio\/} methods. According to the
electronic band-structure calculations reported in the literature
\cite{8,9,10,11,12,13,14}, MgB$_2$ 
can be thought of as a compound made of honeycomb layers of
electronically depleted
Mg and electronically enriched B (Fig.~\ref{fig1}). There are, therefore,
spatially separated layers of Mg and B atoms that can play the roles of 
electron acceptor and donor reservoirs, respectively. Depending on the
degree of flexibility of the donor electrons, one may expect a charge
instability in MgB$_2$ with respect to external perturbations.
Such kind of instability in MgB$_2$ may be responsible for the excitation
of a collective charge transfer mode theoretically predicted in Ref.~\cite{15}. 
This excitation mode involves coherent charge fluctuations between the B and Mg 
layers and, according to Ref.~\cite{15}, may be of relevance for 
understanding the mechanism of high-temperature
superconductivity in MgB$_2$. 

The properties of the ground state of MgB$_2$ that favour charge transfer
processes at low excitation energies may also have an impact on the spectroscopy
of highly excited states. In particular,  the corresponding valence charge 
transfer (CT) excitations  accompanying ionization of Mg core levels may appear
in the spectrum as shake-up (energetically above the intense main line) or even as shake-down 
(below the main line) satellites. 
The intensity and the energy of the satellites are controlled by the strength
of the coupling between the core hole and the respective valence-excited 
core-hole configurations and by their relative energies. To zeroth order
in the interaction with the core hole, the energy of the satellite
relative to the energy of the main
state (the state dominated by the single core hole configuration) is equal to
the energy of the respective valence excitation in the neutral system. In MgB$_2$
the energies of the lowest valence CT excitations are estimated to be 0.5, 2.5 and 5~eV \cite{15}.
We, therefore, expect to find low-lying CT satellites in the
Mg core ionization spectra of this compound. The available experimentally recorded
spectra differ substantially from each other \cite{21,22}

In the present paper we report an accurate {\em ab initio\/} Green's function
study of the ionization of Mg2p electrons in MgB$_6$ and Mg$_2$B$_4$ fragments
of the bulk MgB$_2$ (as indicated in Fig.~\ref{fig1}). In both  clusters the Mg2p spectrum
exhibits markedly intense CT satellites at low energies. We also computed the
ionization of a related new molecular MgB$_6$ cluster which is shown to be a stable species
at its optimal C$_{6v}$ geometry. The general appearance of the spectral features
in this molecular cluster resembles that obtained for the bulk clusters. 

\section{Details of the calculations}

Core ionization spectra were computed using the {\em ab initio\/} Green's function
(GF) method within the algebraic diagrammatic construction scheme (ADC)
in fourth order of the Coulomb interaction (ADC(4)) \cite{16,16a}.
This method takes into account not only all terms through fourth order but also includes 
partial summations of diagrams to infinite order.
The GF ADC(4) method accurately accounts for orbital relaxation. What is
important is that the method maintains a balanced description of orbital relaxation and electron correlation.
 The GF ADC(4) method has proved to yield
a very good description of core ionization \cite{16a,Be,Ni}. 
An integral-driven implementation \cite{17} of the original ADC(4) code
\cite{18} was used, enabling us to handle cases
where the explicit GF matrix would have been too large 
to compute the whole spectrum (the size of the GF matrices
was $\sim$2,400,000 for Mg$_2$B$_4$ and $\sim$1,200,000
for MgB$_6$ clusters).
The Hartree-Fock orbitals
and their energies used as input data for the GF ADC(4) method have been calculated
using the {\em GAMESS-UK\/} package \cite{19} with the 6-311G* atomic basis for B \cite{20}
and Mg \cite{20a}.
In addition, the d function on Mg is selected as 
prescribed in Ref.~\cite{20aa}. 
 The quality of this medium-sized basis set for the GF ADC(4) core-hole
calculations on Mg and Be clusters has been assessed in Refs.~\cite{Be,20b}. 

The nuclear geometry in the MgB$_6$ and Mg$_2$B$_4$ fragments of the MgB$_2$
crystal has been taken from Ref.\cite{1}. The geometry of the MgB$_6$ molecular
cluster has been optimized at the level of  fourth-order M{\o}ller-Plesset many-body
perturbation theory. This results in a planar B$_6$ ring with a B-B
distance of 3.012~a.u. and the Mg atom is situated 3.346~a.u. above the centre of the
ring (C$_{6v}$ symmetry). These distances are slightly different from those of the MgB$_6$
fragment of the MgB$_2$ crystal where the B-B distance is 3.367~a.u. and the Mg atom
is at 3.329~a.u. above the plane of the B ring. For the crystal structure see Fig.~\ref{fig1}.

\section{Results and discussion}

Our computed Mg2p ionization spectra
are shown in Fig.~\ref{fig2} in comparison with the available experimental data.
A key finding of our calculations
is that the electronic structure of all the clusters responds quite strongly to
the creation of the core hole in the Mg2p levels. This strong response is mostly
associated with the B2p$_z\rightarrow$Mg3s,3p charge transfer and manifests itself
in the appearance of satellite structures of considerable intensity, which, in
the spectra of the MgB$_6$ clusters, can be seen at energies of $\sim$~3, 6 and 10~eV (Fig.~\ref{fig2}). 
In the spectrum of the Mg$_2$B$_4$ cluster the B$\rightarrow$Mg charge transfer gives rise to
a prominent satellite peak at $\sim$2~eV, with further
structures again at $\sim$6 and 10~eV.
The satellite peaks have been labelled A, B and C in the
theoretical spectra of Fig.~\ref{fig2}. It is surprising that, as the figure shows,
these intense structures emerge by
accumulation of intensity
of numerous core-hole states.
A general 
feature of all the spectra is that many low-lying individual shake-up and even shake-down satellites appear, some of
them at energies as low as 0.1~eV below the main line, which is defined as the
line of highest intensity. The appearance of  
satellite peak A, especially in the
spectrum of Mg$_2$B$_4$, is in good agreement
with the experimental MgB$_2$ spectrum  of Ref.~\cite{21}, which
also exhibits an intense structure at $\sim$2~eV above the main line.
This differs remarkably from the other experimental
Mg2p spectrum of MgB$_2$ \cite{22}, also shown in Fig.~\ref{fig2}.
The latter shows an utterly different band shape and only
a small hump can be discerned at $\sim$~2~eV above the maximum of the
broad low energy band.

Let us begin our discussion with the Mg$_2$B$_4$ fragment cluster. This cluster has the stoichiometry
of the bulk MgB$_2$ and contains atoms of two adjacent unit cells of the crystal (Fig.\ref{fig1}).
A most intriguing feature of the Mg2p$^{-1}$ spectrum of the cluster is that there is no isolated
main line and actually a dominating main line does not exist at all (Fig.\ref{fig2}).
At low resolution a dominant peak, which we may denote as main peak, is seen in the spectrum, but this
peak is broad and consists of many lines
with rather small individual intensities. The most intense among them
exhibits an intensity as small as 22\% of the total spectral intensity. We encounter here
a breakdown of the main line which is very unusual for core spectra of systems not containing
$d$ and $f$ metal atoms. The mechanism
responsible for this breakdown, as well as for the appearance of the A, B, and C satellite
structures in the spectrum, is a pronounced B2p$_z\rightarrow$Mg3s,3p charge transfer.

In the ground state of the cluster an electron has been transferred from the Mg atoms to the
borons. The creation of the core hole in the Mg2p level initiates a charge transfer 
in the opposite direction. By ionizing the Mg2p level the charge is thus transferred
back from the boron zigzag to the core-ionized Mg sites. This transferred charge screens the initial core
holes giving rise to the appearance of the satellite structures A, B, and C  and to the
breakdown of the main line. 
Each of the structures consists of very many individual states which are strong mixtures
of singly-excited and doubly-excited core-hole electronic configurations,
where in addition to the core hole one or two valence electrons
are excited, respectively. In the
spectrum, the satellites acquire their intensity by borrowing it from the core-hole 
configuration. The intensity borrowed is
controlled by the strength of the coupling between the core-hole and the 
excited configurations. Usually, the intensity acquired by  states that are dominated
by doubly-excited configurations is very small because the coupling of such
configurations to the core hole appears only in second order in 
the Coulomb interaction. By contrast,
singly-excited states  can acquire considerable intensity in the spectrum
because their coupling to the core hole appears already in first order in the Coulomb
interaction. In our case the following picture emerges. The valence doubly-excited states 
appear at energies as low as 
$\sim$~2~eV above the main line and their spectral density is much
higher than that of singly-excited states.  As usual, the intensity is borrowed by the 
singly-excited states, which are CT states in our case. Due to the presence of the
doubly-excited states and their interaction with the singly-excited ones, this intensity
is redistributed among the many states available. Therefore, instead of finding
a few states with appreciable intensities, we find structures A, B and C, which are 
considerably broadened by multiple splittings.
This mechanism
of  broadening may also play a role in forming the shape of the spectrum of solid MgB$_2$.
In the latter case,
the peaks may have rather large widths due to the fact that
the valence excitations, which in the computed Mg$_2$B$_4$ cluster comprise only a limited
number of transitions between  molecular
orbitals, become transitions between electronic bands.

As we mentioned above,
a remarkable feature of the spectrum of Mg$_2$B$_4$ is that peak
A is especially intense and appears at $\sim$2~eV above the main line. Its energy 
and even its intensity relative to
the main peak resemble strikingly those of the structure observed at 2~eV in the experimental
spectrum of Ref.~\cite{21}.

The other cluster model  of the MgB$_2$ crystal, the MgB$_6$ fragment, 
does not have the stoichiometry of MgB$_2$ but contains the main structural motif of the B
layers of the crystal, that is the B$_6$ planar ring. The physics of core-ionization of
this cluster is similar to that found for Mg$_2$B$_4$. Strong B2p$_z\rightarrow$Mg3s,3p
charge transfer processes are responsible for the appearance of the satellite structures
A, B and C in the Mg2p$^{-1}$ spectrum. These structures consist of many lines with small
individual intensities, as  in the case of Mg$_2$B$_4$. 
We
observe also in MgB$_6$ a breakdown of the main line; however, this
breakdown
is not as pronounced as that encountered in the  Mg$_2$B$_4$ cluster. 

The spectrum of the molecular MgB$_6$ cluster resembles to a large extent the
spectrum of the MgB$_6$ fragment of the MgB$_2$ crystal (Fig.~\ref{fig2}).
This is not very surprising because the geometry of the MgB$_6$ molecular
cluster is rather close to that of the MgB$_2$ crystal.
Similarly to the MgB$_6$ and Mg$_2$B$_4$ fragments, the B2p$_z\rightarrow$Mg3s,3p charge transfer
is responsible for the appearance of intense satellite structures
in the spectrum.
The structure of the MgB$_6$ molecular cluster is of
particular interest. According to our calculations the B$_6$ planar ring alone is not stable.
This ring is, however, stabilised by the electronic charge transfer from the Mg atom to the boron ring in the
ground state of MgB$_6$. The similarity of the geometry of this
cluster to that of the crystal indicates that this mechanism may also be of relevance for the stabilisation
of the crystal structure of MgB$_2$.

A general feature of the Mg2p core hole screening in all our clusters is
that the  B2p$_z\rightarrow$Mg3s,3p charge transfer processes are dominant.
Interestingly, the related valence B2p$_z\rightarrow$Mg3s,3p charge transfer
excitations have been shown to be of importance for understanding the mechanism
of high-temperature superconductivity in MgB$_2$ \cite{15}.
Our calculations show that the B2p$_{x,y}\rightarrow$Mg3s,3p charge transfer processes
make only minor contribution to the screening of the core hole. 
According to the band structure calculations, $\sigma$(B2p$_{x,y}$) states,
which are quasi-two-dimensional $\sigma$ states, form the metallic
properties of MgB$_2$ \cite{8,9,10,11,12,13,14}.

According to our calculations, the total intensity borrowed by satellites is $\sim$~60\% in the spectra
of the MgB$_6$ clusters and $\sim$~80\% in the spectrum of Mg$_2$B$_4$. This high satellite to main
line intensity ratio is not typical for light element compounds where satellites usually acquire
$\sim$~20-30\% of the total spectral intensity. In this respect the clusters studied here can rather  
be related to the class of strongly correlated systems. These include also known high-temperature
superconductors. Another feature of the Mg-B clusters, which they share with these superconductors, 
is that the main
mechanism responsible for the appearance of intense satellite structures in the core-level
spectra is a charge transfer mechanism.

It is worthwhile to discuss the experimental Mg2p$^{-1}$ spectra. The band shapes of the 
two experimental spectra available in the literature are very different from each other
(Fig.~\ref{fig2}). According to Ref.~\cite{21}, the surface of the prepared $c$-axis oriented 
thin film of MgB$_2$ grown on an $R$-plane sapphire was initially contaminated with different
Mg oxides, hydroxides and carbonates. The ionization of the Mg2p level in these compounds
contribute to the bulk MgB$_2$ spectrum at energies of $\sim$2~eV. The experimental spectrum
(b) shown in Fig.~\ref{fig2}, however, has been measured for the MgB$_2$ surface cleaned
by non aqueous chemical etching. The authors of Ref.~\cite{21} claimed that the etching yields
the high-quality surface free of surface compounds. They demonstrated that the intensity of
the 2~eV peak in the spectrum indeed decreases considerably after the etching but this peak does not
disappear entirely. 

The authors of Ref.\cite{22} also measured the Mg2p$^{-1}$ spectrum from a thin film of MgB$_2$
grown on an sapphire substrate. In contrast to the spectrum (b) \cite{21} there is only a small
hump observed in the spectrum (a) at 2~eV above the most intense peak
(Fig.~\ref{fig2}). That is not the only difference between the (a) and (b) experimental spectra.
The spectrum (a) reported in Ref.~\cite{22} clearly exhibits an intense line at $\sim$1eV below
the most intense peak. There is no similar structure in the spectrum (b) \cite{21}.
It should be noted that the MgB$_2$ film was grown
by the authors of Ref.~\cite{22} using a method different from that of  Ref.~\cite{21}. The authors
of  Ref.~\cite{22} claimed that their MgB$_2$ surface was initially clean, in contrast to
the film obtained initially in Ref.~\cite{21}. 

It is not clear why the two experimental spectra have so different band shapes. There
can be still impurities in the prepared MgB$_2$ samples or there may be
structural irregularities specific to each of these experiments. We should note,
however, that the spectrum (a) seems to be a superposition of the spectrum (b) and some
other structure contributing to the spectrum at $\sim$~1~eV below the highest peak
(Fig.~\ref{fig2}). The absolute energy of the peak maximum is the same
in both experimental spectra.
Based on our {\em ab initio\/} calculations alone, we cannot identify the factors which
make the experimental spectra measured in Refs.~\cite{21,22} so different. Obviously, this situation
should be clarified experimentally. Our calculations support the experimental
results reported in Ref.~\cite{21}.

We have calculated the spectra of clusters of limited size. Upon increasing the size of the
cluster, one can expect changes in the spectral bandshapes. According to our analysis one of
the effects related to the size of the cluster is the broadening of the satellite structures,
and also some additional structures can appear in the spectra of larger clusters. We argue, however,
that these changes cannot be significant. The main mechanism responsible for the complicated
structure of the spectra of the clusters is the B$\rightarrow$Mg charge transfer. The strength
of this process, and consequently the intensity of the corresponding lines in the spectrum, is
controlled by the overlap between the B2p and Mg3s,3p orbitals. Therefore,
charge transfer from B atoms which lie near the core-ionized Mg site
make the dominant contribution to the screening of the Mg2p core hole,
whereas
charge transfer
from boron atoms far away from the core-ionized Mg atom cannot be strong,
due to a small $\langle{B2p}|Mg3s,3p\rangle$ overlap.
Thus, we do not expect significant modifications of the spectra of larger clusters
compared to the spectra of our relatively small ones. 

The electronic structure of the  MgB$_6$ molecular cluster sheds some
light on the factors stabilising the structure of the MgB$_2$ crystal.
The boron planar honeycomb conformation is stabilised through electron
donation from Mg in the ground state of the cluster.
In the Mg2p core-ionized states this transferred
charge goes back to the Mg site and screens the core hole. 
The strength of this screening in the molecular cluster is similar to that in the cluster
models of the MgB$_2$ crystal and therefore the 
core-ionization spectra are similar as well.

To summarize, our accurate {\em ab initio\/} Green's function calculations reveal strong
many-body charge transfer effects in the Mg2p$^{-1}$ spectra of the MgB$_6$ and Mg$_2$B$_4$
fragments of the MgB$_2$ crystal and in the spectrum of the molecular MgB$_6$ cluster.
Three groups of satellites appear in the spectra at $\sim$~2-3, 6 and 10~eV above the
main line. 
The appearance of the 3~eV peak in our computed spectra and especially
of the peak at 2~eV in the spectrum of Mg$_2$B$_4$ reflects quite closely
the experimental spectrum reported in Ref.~\cite{21}. The two experimental Mg2p$^{-1}$ spectra
of MgB$_2$ available in the literature are very different from each other. This fact, together
with the results of our calculations, call for more elaborate experimental studies on the core-level
spectra of MgB$_2$.

%\begin{acknowledgments}
\section{Acknowledgements}
Financial support by the DFG and by the {\em Programma materiali
innovativi\/} (C.N.R.) is gratefully acknowledged.
%\end{acknowledgments}   

\pagebreak

\begin{figure}
\caption{Crystal structure of MgB$_2$. The atoms of the Mg$_2$B$_4$ and MgB$_6$ fragments 
of the MgB$_2$ crystal studied in the 
present work are indicated with the indices 1 and 2, respectively.}
\label{fig1}
\end{figure}

\begin{figure}
\caption{The computed Mg2p$^{-1}$ ionization spectra of the MgB$_6$, Mg$_2$B$_4$ fragments
of the MgB$_2$ crystal and of the molecular cluster MgB$_6$ in comparison with the experimental data (a) and (b)
reported in Refs.~\cite{22} and \cite{21}, respectively. 
The relative positions of the experimental spectra have been obtained using their original
absolute ionization energy scales.
The theoretical spectra were aligned with respect
to the position of the peak with the maximal intensity. 
The vertical lines represent the relative ionization 
energies and intensities of the computed discrete spectral lines.
The band shapes have been obtained
by Gaussian broadening of these discrete lines with a half width of 0.8~eV.}
\label{fig2}
\end{figure}

\end{document}